# Emergent energy scales in magnonic systems with relative motion

**Daigo Oue**[1,2]

[1] RIKEN Centre for Advanced Photonics, RIKEN, Saitama 351-0198, Japan
[2] The Blackett Laboratory, Imperial College London, London SW7 2AZ, United Kingdom

E-mail: daigo.oue@gmail.com

**Abstract**

Relative motion between interacting systems can generate emergent energy scales that are absent in isolated systems. While uniform motion can be eliminated by a Galilean transformation, relative motion between interacting systems generally cannot. By coupling to an excitation's spatial structure, relative motion generates a Doppler frequency determined by its wavevector and the relative velocity, providing a mechanism for driving nonequilibrium phenomena.

In this tutorial, we illustrate these ideas using magnonic systems as a concrete platform. We first discuss motion-induced magnon transport between relatively moving ferromagnets, in which the Doppler frequency serves as an effective nonequilibrium bias in the perturbative regime. This mechanism produces magnon currents even without conventional driving forces such as temperature gradients or chemical potential differences.

We then introduce motion-induced parametric instabilities. When the emergent scale becomes sufficiently large to resonantly create magnon pairs, the perturbative description breaks down, and the magnonic vacuum becomes unstable. This instability occurs above a critical velocity threshold and leads to spontaneous magnon-pair creation.

Connections to related phenomena, including quantum friction, Cherenkov emission, and Zel'dovich superradiance, are also highlighted. The concept of an emergent energy scale provides a unifying framework for understanding transport phenomena and instabilities in quantum systems with relative motion.

**Glossary**

*Magnon, Doppler effect, Transport phenomena, Instability, Galilean transformation, PT symmetry*

## Introduction

Relative motion is ubiquitous in nonequilibrium physics, appearing in a wide range of systems from sheared interfaces to counter-flowing plasmas. Such relative motion gives rise to a wide variety of nonequilibrium phenomena and has attracted considerable interest across many branches of physics. At first sight, this nonequilibrium nature may seem counterintuitive, since uniform motion of an isolated system has no observable consequence: it can always be eliminated by changing the reference frame and is therefore indistinguishable from equilibrium.

The situation changes qualitatively when multiple interacting systems move relative to one another. Although each subsystem may undergo uniform motion, no single inertial frame can simultaneously bring all interacting subsystems to rest. In this sense, relative motion is piecewise uniform but not globally uniform. It may therefore be regarded as a discontinuous velocity field, providing a fundamentally nonequilibrium platform for interacting systems.

Indeed, relative motion induces a broad range of nonequilibrium phenomena, most notably transport processes and dynamical instabilities. For example, sheared interfaces experience friction through momentum transport across the interface [1], whereas counter-flowing plasmas exhibit dynamical instabilities associated with the softening of collective modes [2]. Despite their different manifestations, both ultimately stem from the same piecewise but globally non-uniform relative motion. One of the central messages of this tutorial is that both can be understood in terms of an emergent energy scale arising from relative motion.

In this tutorial, we take magnonic systems as a simple and instructive platform to illustrate these ideas. Magnons are not only collective spin excitations central to spin transport but have also emerged as controllable bosonic degrees of freedom in hybrid quantum systems, where their populations, interactions, and quantum correlations can be manipulated and probed [3]–[7]. Such recent developments in magnonic platforms further highlight their versatility for exploring coupled bosonic dynamics. In particular, we show how relative motion, via an emergent energy scale, gives rise to motion-induced transport phenomena [8][9] and dynamical instabilities [10] in interacting ferromagnets. The underlying physical picture is applicable well beyond magnonics, and we briefly discuss connections to related phenomena in other fields, including quantum friction [1], Cherenkov emission [11], and superradiant amplification [12].

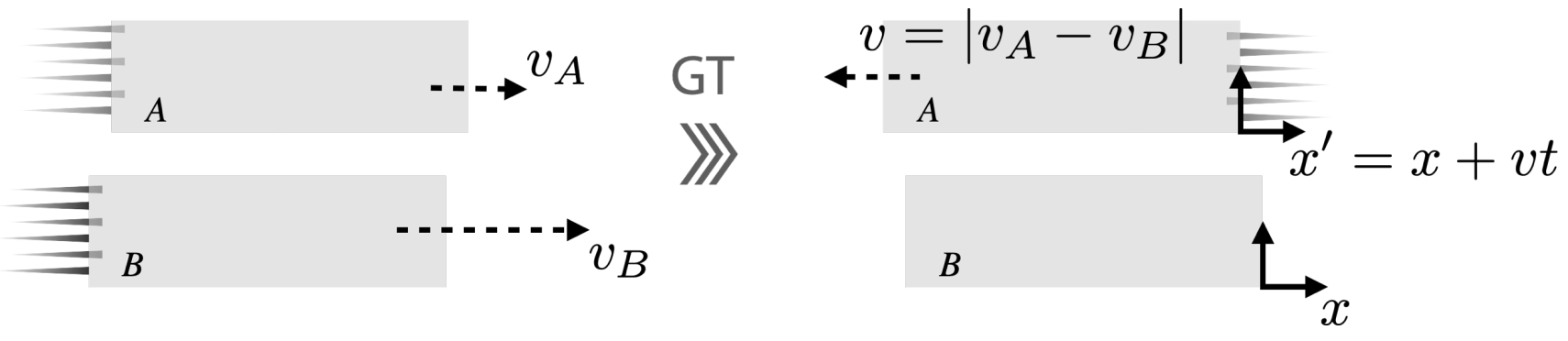

**Figure 1.** Relatively moving subsystems A and B and the reference frames related by the Galilean transformation (GT). Subsystem B is at rest in the reference frame used in the main text.

## Emergent energy scales induced by relative motion

Relative motion alone does not define an emergent frequency scale. Indeed, the relative speed $v = |\boldsymbol{v}|$ has the dimension of velocity and therefore cannot be directly compared with the characteristic frequencies of a physical system. To generate a frequency scale, the velocity must be combined with a characteristic length scale $\ell$ of the system. Dimensional analysis then immediately suggests the emergent frequency scale $\Delta = 2\pi v/\ell$, which represents the simplest frequency scale associated with relative motion.

The characteristic length should not, in general, be regarded as a single geometric length chosen independently of the excitation spectrum. Rather, it represents the spatial scale of the excitation that couples to the relative motion. For an excitation with wavenumber $k$, its wavelength $\ell = 2\pi/k$ gives $\Delta = kv$, or more generally $\boldsymbol{k} \cdot \boldsymbol{v}$ when the propagation direction is considered. In practical systems, geometric and material length scales, such as the film thickness or intrinsic magnetic length scales, influence the excitation spectrum and thereby determine which wavevectors are physically relevant (we return to this point in the fourth section). This identification follows directly from the transformation of a finite-wavevector excitation: Consider two interacting systems in relative motion and choose the inertial frame in which one is at rest, as illustrated in Fig. 1. In this frame, the other subsystem moves with relative velocity $\boldsymbol{v}$. An elementary excitation on the moving subsystem with a dispersion relation $\omega_{\boldsymbol{k}}$ has the phase factor $\exp[i(\boldsymbol{k} \cdot \boldsymbol{x}' - \omega_{\boldsymbol{k}} t)]$, where the coordinates satisfy $\boldsymbol{x}' = \boldsymbol{x} + \boldsymbol{v}t$. Expressing the phase in terms of the stationary coordinate gives $\boldsymbol{k} \cdot \boldsymbol{x}' - \omega_{\boldsymbol{k}} t = \boldsymbol{k} \cdot \boldsymbol{x} - (\omega_{\boldsymbol{k}} - \boldsymbol{k} \cdot \boldsymbol{v})t$, showing that the excitation is observed with the Doppler-shifted frequency $\omega_{\boldsymbol{k}} \to \omega_{\boldsymbol{k}} - \boldsymbol{k} \cdot \boldsymbol{v}$.

The emergent frequency scale predicted by the dimensional argument is therefore identified as $\Delta = k_v v$ with the projected wavenumber $k_v = \boldsymbol{k} \cdot \boldsymbol{v}/|\boldsymbol{v}|$. This is nothing but the Doppler shift of a finite-wavevector excitation. This immediately explains why the uniform mode ($\boldsymbol{k} = 0$) is unaffected by relative motion, whereas finite-$\boldsymbol{k}$ excitations experience a frequency shift determined by the projection of the relative velocity onto the wavevector. The above discussion is based on a Galilean transformation and therefore applies in the non-relativistic regime. For relativistic velocities, the frequency shift should instead be obtained from the Lorentz transformation.

Once interactions between the two subsystems are introduced, the emergent frequency scale should appear in the interaction. In general, we can consider two types of resonance conditions: difference-frequency and sum-frequency conditions,

$$\omega_b - \omega_a = \Delta, \qquad \omega_b + \omega_a = \Delta. \tag{1}$$

Which resonance condition becomes relevant depends on the interaction channel. Two fundamental excitations (bosonic modes on each subsystem) are described by the creation and annihilation operators $a_k^\dagger$, $a_k$ and $b_k^\dagger$, $b_k$. The interaction Hamiltonian can therefore be classified according to whether it exchanges excitations between the two subsystems or simultaneously creates (or annihilates) excitation pairs. The most general bilinear interaction between two bosonic modes is constructed from four operator products, $a_{\boldsymbol{k}} b_{-\boldsymbol{k}}$, $a_{\boldsymbol{k}} b_{\boldsymbol{k}}^\dagger$, $a_{\boldsymbol{k}}^\dagger b_{\boldsymbol{k}}$, $a_{\boldsymbol{k}}^\dagger b_{-\boldsymbol{k}}^\dagger$, which, due to the Hermiticity, naturally combine into two independent channels,

$$H_{\mathrm{ex}} = \sum_{\boldsymbol{k}} \{g_{\mathrm{ex}}(\boldsymbol{k})\}(a_{\boldsymbol{k}}^\dagger b_{\boldsymbol{k}} + a_{\boldsymbol{k}} b_{\boldsymbol{k}}^\dagger), \qquad H_{\mathrm{nc}} = \sum_{\boldsymbol{k}} \{g_{\mathrm{nc}}(\boldsymbol{k})\}(a_{\boldsymbol{k}}^\dagger b_{-\boldsymbol{k}}^\dagger + a_{\boldsymbol{k}} b_{-\boldsymbol{k}}), \tag{2}$$

where $g_{\mathrm{ex}}$ and $g_{\mathrm{nc}}$ are interaction strengths associated with the two interaction channels. The first channel conserves the total number of excitations by exchanging an excitation between the two subsystems, whereas the second channel simultaneously creates or annihilates a pair of excitations (see Fig. 2).

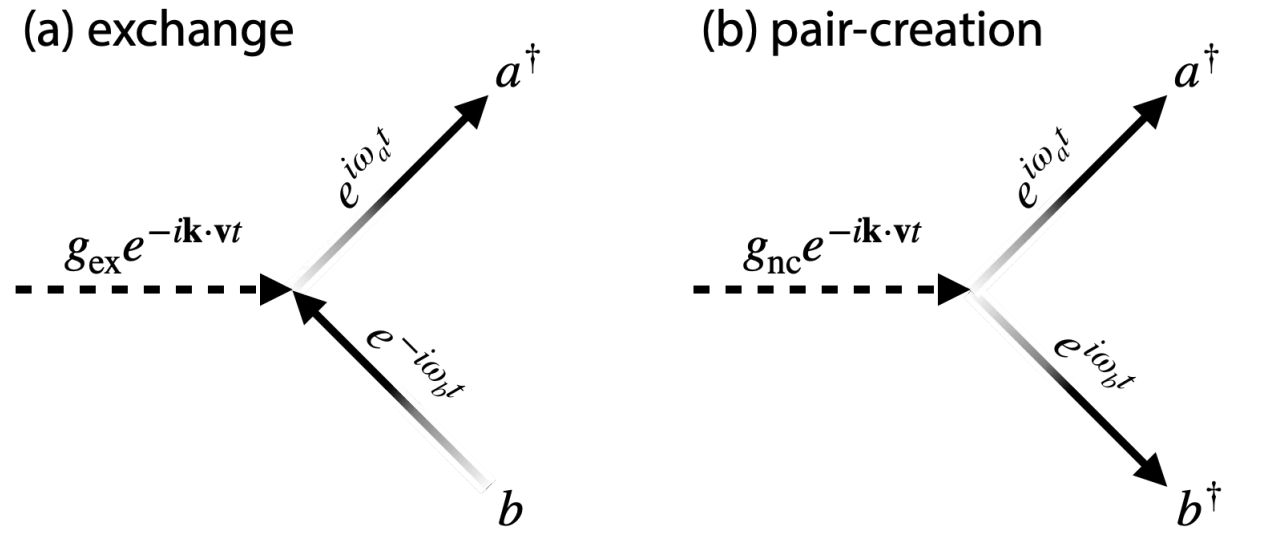

**Figure 2.** Two bilinear interaction channels between the subsystems: (a) excitation-exchange channel and (b) excitation-pair-creation channel.

Suppose that the excitation represented by $a$ ($b$) belongs to the moving (stationary) subsystem. The Doppler shift derived above then directly determines the free evolution of the two excitations. In the reference frame in which subsystem $b$ is at rest, the excitation frequency of the moving subsystem is shifted as $\omega_a \to \omega_a - \boldsymbol{k} \cdot \boldsymbol{v}$, whereas that of the stationary subsystem remains $\omega_b$. Here, the dependence on the wavevector $\boldsymbol{k}$ in the dispersion relation is left implicit. We further assume that the relative velocity is given by $\boldsymbol{v} = v\boldsymbol{u}_x$ with $\boldsymbol{u}_x$ the unit vector in the $x$ direction. The corresponding unperturbed Hamiltonian is therefore given by

$$H_0 = \sum_k \left\{(\omega_a - k_x v) a_{\boldsymbol{k}}^\dagger a_{\boldsymbol{k}} + \omega_b b_{\boldsymbol{k}}^\dagger b_{\boldsymbol{k}}\right\}. \tag{3}$$

A microscopic derivation of the corresponding magnon Hamiltonian is provided in the Supplementary Information; see also Ref. [9] for further details. Going to the interaction picture [e.g. $O \to O(t) = e^{+iH_0 t} O e^{-iH_0 t}$], we can write

$$H_{\mathrm{ex}}(t) = \sum_k g_{\mathrm{nc}}(\boldsymbol{k}) e^{i(\omega_a - \omega_b - k_x v)t} a_{\boldsymbol{k}}^\dagger b_{\boldsymbol{k}} + \mathrm{h.c.}, \qquad H_{\mathrm{nc}}(t) = \sum_k g_{\mathrm{nc}}(\boldsymbol{k}) e^{i(\omega_a + \omega_b - k_x v)t} a_{\boldsymbol{k}}^\dagger b_{-\boldsymbol{k}}^\dagger + \mathrm{h.c.} \tag{4}$$

The origin of the two resonance conditions (1) now becomes apparent: The phase factor associated with each channel contains exactly the resonance conditions anticipated above. Resonance occurs when the corresponding phase factor becomes stationary; therefore, the exchange-conserving (non-conserving) channel is resonantly enhanced under the difference-frequency (sum-frequency) condition.

The following two sections discuss how the difference-frequency resonance gives rise to transport phenomena, whereas the sum-frequency resonance leads to dynamical instabilities.

**Transport driven by emergent energy**

When the emergent frequency scale is sufficiently small, the excitation-conserving interaction channel becomes the relevant one (so we consider $H = H_0 + H_{\mathrm{ex}}$ in this section). This interaction is resonantly enhanced under the difference-frequency condition and serves as an effective nonequilibrium bias that drives transport. It transfers an excitation from one subsystem to the other while conserving the total number of excitations, and its physical consequence is therefore naturally a transport phenomenon. In conventional transport, temperature or chemical-potential differences typically create nonequilibrium. Here, relative motion provides an alternative source of nonequilibrium through the emergent frequency scale generated by the Doppler shift. Consequently, we can understand the transport perturbatively as a response to this effective nonequilibrium bias.

For example, consider two identical ferromagnetic insulators: one moving at a constant velocity while the other remains at rest. The magnon dispersion relation is given by

$$\omega_{a,b} = Dk^2 + \omega_0, \tag{5}$$

where $D$ is the spin-wave stiffness and $\omega_0$ denotes the Zeeman energy. Note that we consider only the $x$ direction for simplicity, as it is the direction that is most significantly affected by the Doppler effect $k_x v$. Applying time-dependent perturbation theory with respect to the exchange interaction $H_{\mathrm{ex}}$, the spin transfer carried by magnons between the two magnets can be evaluated analytically [8][9], yielding

$$I_{\mathrm{s}} = 4\int |\hbar g_{\mathrm{ex}}(k)|^2 \Im\left(\frac{2S_0/\hbar}{\omega-(\omega_a-kv)+i\Gamma}\right)\Im\left(\frac{2S_0/\hbar}{\omega-\omega_b+i\Gamma}\right)[n_{\mathrm{B}}(\omega_b)-n_{\mathrm{B}}(\omega_a-kv)]\,\mathrm{d}k\,\mathrm{d}\omega. \tag{6}$$

Here, $\Gamma$ phenomenologically accounts for the finite magnon lifetime and associated spectral broadening, $S_0$ is the magnitude of the localised spin, and $n_{\mathrm{B}}$ is the Bose distribution. Equation (6) describes transport through the overlap of the broadened magnon spectra and the motion-induced population imbalance between the two subsystems. The detailed derivation can be found in Refs. [8][9].

An interesting consequence is that the spin transfer is proportional to the second order of the relative velocity. This follows because the Doppler shift always appears through the combination $kv$. For the reciprocal and inversion-symmetric configuration considered here, the magnon dispersion and the interaction are even under momentum reversal, $\omega_k = \omega_{-k}$ and $|g_{\mathrm{ex}}(k)|^2 = |g_{\mathrm{ex}}(-k)|^2$. The contribution linear in $v$ is therefore odd in $k$ and vanishes upon integration, even when the interaction strength is momentum-dependent. The leading contribution is consequently proportional to $v^2$. A linear-in-$v$ contribution may arise when this momentum-reversal symmetry is broken. This is consistent with the intuitive picture that the transport is driven by the kinetic energy associated with the relative motion, which is itself proportional to $v^2$.

**Instabilities driven by emergent energy**

When the emergent frequency scale is sufficiently large, the excitation-nonconserving interaction channel becomes the relevant one (so we consider $H = H_0 + H_{\mathrm{nc}}$ in this section). This interaction is resonantly enhanced under the sum-frequency resonance condition and simultaneously creates or annihilates excitation pairs. Unlike the exchange channel discussed in the previous section, the consequence here is not transport between two subsystems but the amplification of collective excitations. In this regime, the emergent frequency scale no longer acts merely as an effective nonequilibrium bias. Instead, it provides the energy required for pair creation through an effective parametric drive. Once pair creation becomes resonant, the excitation population grows exponentially, eventually leading to a dynamical instability.

Before explicitly treating the interaction, it is useful to estimate the threshold using the resonance condition. For two identical ferromagnets with the magnon dispersion relation, $\omega_{a,b} = Dk^2 + \omega_0$, the sum-frequency resonance condition, $\omega_a + \omega_b - vk = 0$, gives

$$2Dk^2 - vk + 2\omega_0 = 0. \quad (7)$$

This equation has real solutions when $v^2 - 16D\omega_0 > 0$, providing the threshold estimate $v_c \simeq 4\sqrt{D\omega_0}$. This identifies the regime in which magnon-pair creation becomes allowed. It also illustrates how the excitation spectrum selects the characteristic length introduced in the second section. The velocity required to satisfy the resonance is $v = 2\omega_k/k = 2Dk + 2\omega_0/k$, which is minimised at $k_{\min} = \sqrt{\omega_0/D}$ (The corresponding wavelength is $\ell_{\min} = 2\pi\sqrt{D/\omega_0}$). Since $\omega_{k_{\min}} = 2\omega_0$, the resonance condition $k_{\min}v_c = 2\omega_{k_{\min}}$ gives $v_c = 4\sqrt{D\omega_0}$.

To determine the actual stability of the interacting system, we consider the equations of motion for the magnon creation and annihilation operators. Collecting the operators into a vector, the Heisenberg equations can be written as [10]

$$i\frac{\partial}{\partial t}\begin{pmatrix} a_k \\ b_{-k} \\ a_k^\dagger \\ b_{-k}^\dagger \end{pmatrix} = \begin{pmatrix} \omega_a - kv & 0 & 0 & g_{\mathrm{nc}} \\ 0 & \omega_b & g_{\mathrm{nc}} & 0 \\ 0 & -g_{\mathrm{nc}} & -\omega_a + kv & 0 \\ -g_{\mathrm{nc}} & 0 & 0 & -\omega_b \end{pmatrix}\begin{pmatrix} a_k \\ b_{-k} \\ a_k^\dagger \\ b_{-k}^\dagger \end{pmatrix} \quad (8)$$

The stability is determined by the eigenvalues of the Liouvillian [the 4-by-4 matrix in Eq. (8)], which are

$$\pm\frac{1}{2}\left(kv \pm \sqrt{(\omega_a + \omega_b - kv)^2 - 4g_{\mathrm{nc}}^2}\right). \quad (9)$$

When the square root becomes imaginary, one eigenvalue acquires a positive imaginary part, and the corresponding mode grows exponentially in time. The growth rate is therefore determined by

$$\Gamma_k = \frac{1}{2}\sqrt{4g_{\mathrm{nc}}^2 - (\omega_a + \omega_b - kv)^2} \quad (10)$$

in the parameter region where the quantity inside the square root is positive. For the parabolic magnon dispersion, this condition can be written as $(2Dk^2 - vk + 2\omega_0)^2 < 4g_{\mathrm{nc}}^2$. Equivalently, the instability occurs when $v > v_c = 4\sqrt{D(\omega_0 - g_{\mathrm{nc}})}$. Within the quadratic model, this result is nonperturbative in $g_{\mathrm{nc}}$; the weak-interaction assumption ($g_{\mathrm{nc}} \ll \omega_0$) is required only to recover the resonance-based estimate $v_c \simeq 4\sqrt{D\omega_0}$. The expression for the critical velocity should be understood within the stable regime $g_{\mathrm{nc}} < \omega_0$. As $g_{\mathrm{nc}}$ approaches $\omega_0$ from below, the system approaches an intrinsic instability already present at zero relative velocity, and the motion-induced threshold continuously decreases to zero. For $g_{\mathrm{nc}} > \omega_0$, the system is already unstable at $v = 0$, so the formally imaginary value of $v_c$ should not be interpreted as a physical critical velocity; rather, a motion-induced threshold is no longer defined. The detailed derivation and stability analysis can be found in Ref. [10].

The present treatment is based on the spin-wave approximation. The instability arises within the resulting quadratic Hamiltonian through the pair creation. At sufficient magnon populations, magnon–magnon interactions become relevant and saturate the instability, while the emergence of nonlinear effects after the instability sets in does not invalidate the linear stability analysis used to determine its onset.

Although the Liouvillian above already allows us to determine the instability, it is instructive to relate its structure to $PT$-symmetric systems. The same eigenvalue problem can be rewritten, by rearranging the components and applying a unitary similarity transformation, in the form (See the Supplementary ),

$$\begin{pmatrix} \omega_a - kv & 0 & 0 & g_{\mathrm{nc}} \\ 0 & \omega_b & g_{\mathrm{nc}} & 0 \\ 0 & -g_{\mathrm{nc}} & -\omega_a + kv & 0 \\ -g_{\mathrm{nc}} & 0 & 0 & -\omega_b \end{pmatrix} \mapsto \begin{pmatrix} \delta_k + ig_{\mathrm{nc}} & \Omega_k & 0 & 0 \\ \Omega_k & \delta_k - ig_{\mathrm{nc}} & 0 & 0 \\ 0 & 0 & -\delta_k - ig_{\mathrm{nc}} & -\Omega_k \\ 0 & 0 & -\Omega_k & -\delta_k + ig_{\mathrm{nc}} \end{pmatrix} \quad (11)$$

where we defined $\delta_k = (\omega_a - kv - \omega_b)/2$ and $\Omega_k = (\omega_a - kv + \omega_b)/2$ . This matrix has the same mathematical structure as the balanced-gain-and-loss two-mode matrix familiar from $PT$-symmetric systems [13]. Importantly, this is a unitary similarity transformation of the Liouvillian matrix, not a Bogoliubov transformation defining new bosonic quasiparticles; it therefore preserves the Liouvillian eigenvalues and requires no additional positive-norm condition. The eigenvalues become complex for $|\Omega_k| < g_{\mathrm{nc}}$ (i.e., $(2\omega_k - kv)^2 < 4g_{\mathrm{nc}}^2$ for $\omega_{a,b} = \omega_k$), precisely reproducing the instability condition obtained from Eq. (8), while making its interpretation as spontaneous $PT$-symmetry breaking transparent. Below the threshold, the eigenvalues remain real; above the threshold, they become complex, and the positive imaginary part gives the exponential growth discussed above.

The magnonic instability considered here provides a particularly simple example of a broader class of Doppler-induced phenomena, whose physical consequences depend on the excitation spectrum and the interaction channel, as discussed in the following section.

**Connections to other motion-induced phenomena**

The viewpoint developed in this Tutorial is not specific to magnonic systems but applies more broadly to interacting bosonic systems that support collective excitations, including phonons, plasmons, photons, and polaritons. The Doppler shift provides a common kinematic ingredient, while the dispersion of excitations determines physical consequences.

The role of the excitation spectrum becomes clear when comparing different dispersions. For a gapless linear dispersion, the Doppler condition typically reduces to a direct velocity comparison as in Cherenkov radiation. By contrast, the gapped dispersion considered above introduces intrinsic frequency and wavevector scales. Their competition with the Doppler effect yields a finite characteristic velocity proportional to $\sqrt{D\omega_0}$.

The interaction channel provides a second distinction. In the magnonic example discussed in the third section, an excitation-conserving interaction combines with the Doppler shift to produce a difference-frequency resonance and spin transport. In the fourth section, an excitation-nonconserving interaction instead produces a sum-frequency resonance and magnon-pair creation. The latter has a particularly close analogue in quantum friction: although the relevant excitations and their dispersions differ, the underlying process can likewise be understood in terms of motion enabling the pair creation through a nonconserving interaction [14]–[19].

Other motion-induced phenomena illustrate different consequences of the same Doppler principle. Cherenkov emission converts translational motion into coherent radiation [20]–[22], while drifting electrons transfer energy to propagating acoustic waves in the current-induced acoustic amplification [23][24]. Zel'dovich superradiance [12] provides a rotational counterpart governed by the rotational Doppler shift. Table 1 summarises these examples. The pair-creation viewpoint also connects to the dynamical Casimir effect, where a time-dependent boundary or material parameter acts as a parametric drive and generates correlated photon pairs from vacuum fluctuations [25]–[28].

| Phenomenon | Excitation | Motion |
|---|---|---|
| Magnon transport | Magnon | Translational |
| Magnon instability | Magnon | Translational |
| Quantum friction | Photon | Translational |
| Cherenkov emission | Photon | Translational |
| Drift-induced acoustic amplification | Phonon | Translational |
| Zel'dovich superradiance | Photon | Rotational |

**Table 1.** Motion-induced transport and instabilities.

What is universal is the Doppler scale generated by motion; what is system-specific is how this scale competes with the excitation spectrum and enters the relevant interaction channel. This distinction determines whether relative motion produces transport, coherent radiation, amplification, or pair-creation instability.

## Summary

In this tutorial, we have shown how relative motion between interacting magnetic systems gives an emergent frequency scale through the Doppler effect. The scale depends on how the relative motion probes the spatial structure of excitations, providing a common viewpoint for understanding transport and dynamical instabilities.

In the perturbative regime, the Doppler shift acts as an effective nonequilibrium bias, driving magnon transport even without conventional driving mechanisms such as temperature gradients or chemical potential differences. At larger Doppler shifts, magnon-pair creation becomes resonant, and the system undergoes a dynamical instability above a critical velocity.

These two regimes illustrate how the same motion-induced energy scale can produce qualitatively different phenomena depending on the excitation spectrum and interaction channel. Connections to quantum friction, Cherenkov emission, acoustic amplification, and superradiance further demonstrate the broader relevance of this viewpoint to motion-induced phenomena.

## Acknowledgements

D.O. is supported by the RIKEN Special Postdoctoral Researcher Program. This work was supported by JSPS KAKENHI Grant Numbers JP26K17783 and 26H00686 and by RIKEN Incentive Research Project.

## References

[1] Dedkov GV and Kyasov AA 2017 *Physics-Uspekhi* **60** 559
[2] Sydoruk O, Shamonina E, Kalinin V and Solymar L 2010 *Phys. Plasmas* **17** 102103
[3] Lachance-Quirion D, Tabuchi Y, Gloppe A, Usami K and Nakamura Y 2019 *Appl. Phys. Express* **12** 070101
[4] Yuan HY, Cao Y, Kamra A, Duine RA and Yan P 2022 *Phys. Rep.* **965** 1
[5] Rani S, Cao X, Baptista AE, Hoffmann A and Pfaff W 2025 *Phys. Rev. Applied* **23** 064032
[6] Fan Z, Zuo X, Qian H and Li J 2023 *Photonics* **10** 739
[7] Du HJ, Ma XY and Guo JL 2024 *Quantum Inf. Process.* **23** 198
[8] Oue D and Matsuo M 2022 *Phys. Rev. B* **105** L020302
[9] Oue D and Matsuo M 2024 *physica status solidi (b)* **261** 2300469
[10] Uto T and Oue D 2025 *Phys. Rev. B* **112** 024305
[11] Ginzburg VL 1996 *Physics-Uspekhi* **39** 973
[12] Zel'Dovich YB 1971 *Sov. JETP Lett.* **14** 180
[13] El-Ganainy R, Makris KG, Khajavikhan M, Musslimani ZH, Rotter S and Christodoulides DN 2018 *Nat. Phys.* **14** 11
[14] Pendry JB 1997 *J. Phys.: Condens. Matter* **9** 10301
[15] Volokitin A and Persson BN 2007 *Rev. Mod. Phys.* **79** 1291
[16] Silveirinha MG 2014 *New J. Phys.* **16** 063011
[17] Brevik I, Shapiro B and Silveirinha MG 2022 *Int. J. Mod. Phys. A* **37** 2241012
[18] Oue D, Pendry JB, and Silveirinha MG 2024 *Phys. Rev. Research* **6** 043074
[19] Oue D, Shapiro B and Silveirinha MG 2025 *Phys. Rev. B* **111** 075403
[20] Maghrebi MF, Golestanian R, and Kardar M 2013 *Phys. Rev. A* **88** 042509
[21] Roques-Carmes C, Rivera N, Joannopoulos JD, Soljačić M, and Kaminer I 2018 *Phys. Rev. X* **8** 041013
[22] Fares H and Almokhtar M 2019 *Phys. Lett. A* **383** 1005
[23] Spector HN 1962 *Phys. Rev.* **127** 1084
[24] Barajas-Aguilar AH, Zion J, Sequeira I, Barabas AZ, Taniguchi T, Watanabe K, Barrett EB, Scaffidi T and Sanchez-Yamagishi JD 2024 *Nature Communications* **15** 2550
[25] Moore GT 1970 *J. Math. Phys.* **11** 2679
[26] Dodonov VV 2010 *Phys. Scr.* **82** 038105
[27] Wilson CM, Johansson G, Pourkabirian A, Simoen M, Johansson JR, Duty T, Nori F and Delsing P 2011 *Nature* **479** 376
[28] Lähteenmäki P, Paraoanu GS, Hassel J and Hakonen PJ 2013 *Proc. Natl. Acad. Sci. USA*. **110** 4234

# Supplementary Material

**Emergent energy scales in magnonic systems with relative motion**

*Daigo Oue*

## Supplementary Note 1. From the Heisenberg ferromagnet to the magnon Hamiltonian

This note gives a self-contained derivation of the free magnon Hamiltonian used in the main text. We consider a nearest-neighbour Heisenberg ferromagnet in a magnetic field and work within the linearised spin-wave approximation. We set $\hbar = 1$, as in the main text.

$$H = -J \sum_{\langle i,j \rangle} S_i \cdot S_j - h \sum_i S_i^z \tag{S1}$$

where $J > 0$ is the ferromagnetic exchange coupling and h denotes the Zeeman energy. Introducing Holstein–Primakoff bosons $b_i$, the spin operators are

$$S_i^z = S_0 - b_i^\dagger b_i, \qquad S_i^+ = \sqrt{2S_0 - b_i^\dagger b_i} b_i, \quad S_i^- = b_i^\dagger \sqrt{2S_0 - b_i^\dagger b_i}. \tag{S2}$$

For small deviations from the fully polarised state, $\langle b_i^\dagger b_i \rangle \ll 2S_0$, we retain terms up to quadratic order:

$$S_i^+ \simeq \sqrt{2S_0} b_i, \quad S_i^- \simeq \sqrt{2S_0} b_i^\dagger, \quad S_i^z = S_0 - b_i^\dagger b_i. \tag{S3}$$

Substitution into Eq. (S1), followed by omission of the constant ground-state energy, gives

$$H \simeq h \sum_i b_i^\dagger b_i - J S_0 \sum_{\langle i,j \rangle} (b_i^\dagger b_j + b_j^\dagger b_i), \tag{S4}$$

where $z$ is the coordination number. With the Fourier transformation,

$$b_i = \sum_{\boldsymbol{k}} b_{\boldsymbol{k}} \exp(i\boldsymbol{k} \cdot \boldsymbol{R}_i), \tag{S5}$$

the quadratic Hamiltonian becomes diagonal in wavevector:

$$H \simeq \sum_{\boldsymbol{k}} \omega_{\boldsymbol{k}} b_{\boldsymbol{k}}^\dagger b_{\boldsymbol{k}}, \qquad \omega_{\boldsymbol{k}} = h + J S_0 \sum_{\boldsymbol{\delta}} \exp(i\boldsymbol{k} \cdot \boldsymbol{\delta}), \tag{S6}$$

where the sum runs over the nearest-neighbour vectors $\boldsymbol{\delta}$. Near the ferromagnetic band minimum, the dispersion is parabolic and can be written in the form used in the main text,

$$\omega_k \simeq \omega_0 + Dk^2, \tag{S7}$$

with $\omega_0$ the gap (the Zeeman frequency in the minimal model) and $D$ the spin-wave stiffness. For two uncoupled magnetic subsystems A and B, we therefore write

$$H_0 = \sum_k \left( \omega_a a_{\boldsymbol{k}}^\dagger a_{\boldsymbol{k}} + \omega_b b_{\boldsymbol{k}}^\dagger b_{\boldsymbol{k}} \right). \tag{S8}$$

We now choose the inertial frame in which subsystem B is at rest and let subsystem A move with velocity $\boldsymbol{v}$. As shown in the main text, a finite-wavevector excitation on the moving subsystem transforms according to

$$\omega_a \to \omega_a - \boldsymbol{k} \cdot \boldsymbol{v}. \tag{S9}$$

The free Hamiltonian in this frame is consequently

$$H_0 = \sum_k \left( (\omega_a - \boldsymbol{k} \cdot \boldsymbol{v}) a_{\boldsymbol{k}}^\dagger a_{\boldsymbol{k}} + \omega_b b_{\boldsymbol{k}}^\dagger b_{\boldsymbol{k}} \right). \tag{S10}$$

For $\boldsymbol{v} = v\boldsymbol{u}_x$, Eq. (S11) reduces to the unperturbed Hamiltonian used in the main text, with $\boldsymbol{k} \cdot \boldsymbol{v} = k_x v$. The derivation therefore requires (i) the linearised spin-wave approximation and (ii) the non-relativistic Galilean transformation used to obtain the Doppler shift.

## Supplementary Note 2. Reduction of the Liouvillian to a PT-symmetric two-mode form

Here we give the matrix steps connecting the Liouvillian eigenvalue problem to the balanced-gain-and-loss two-mode form familiar from $PT$-symmetric systems. This is a similarity transformation of the matrix governing the Heisenberg equations; the transformed components are not interpreted as new canonical bosonic operators.

The Liouvillian in the basis $\Psi = \left(a_k, b_{-k}, a_k^\dagger, b_{-k}^\dagger\right)^\top$ is

$$L_k = \begin{pmatrix} \omega_a - kv & 0 & 0 & g_{\text{nc}} \\ 0 & \omega_b & g_{\text{nc}} & 0 \\ 0 & -g_{\text{nc}} & -\omega_a + kv & 0 \\ -g_{\text{nc}} & 0 & 0 & -\omega_b \end{pmatrix}. \tag{S11}$$

Reordering the components as $\Psi' = \left(a_k, b_{-k}^\dagger, b_{-k}, a_k^\dagger\right)^\top$ block-diagonalises the matrix:

$$PL_kP^\dagger = \begin{pmatrix} M_k & 0 \\ 0 & -M_k \end{pmatrix}, \tag{S12}$$

$$M_k = \begin{pmatrix} \omega_a - kv & g_{\text{nc}} \\ -g_{\text{nc}} & -\omega_b \end{pmatrix}, \tag{S13}$$

Introduce the unitary matrix

$$U = \frac{1}{\sqrt{2}}\begin{pmatrix} 1 & -i \\ 1 & i \end{pmatrix}, \qquad UU^\dagger = I. \tag{S14}$$

Direct multiplication gives

$$UM_kU^\dagger = \begin{pmatrix} \delta_k + ig_{\text{nc}} & \Omega_k \\ \Omega_k & \delta_k - ig_{\text{nc}} \end{pmatrix}, \tag{S15}$$

$$\delta_k = \frac{\omega_a - kv - \omega_b}{2}, \qquad \Omega_k = \frac{\omega_a - kv + \omega_b}{2}. \tag{S16}$$